\documentclass[aps,showpacs,superscriptaddress,twocolumn,10pt,nofootinbib]{revtex4-1}

\usepackage{dcolumn}

\usepackage{graphicx}  
\usepackage{dcolumn}   
\usepackage{bm}        
\usepackage{amsthm,amssymb}   
\usepackage{amsmath}
\usepackage{amssymb}
\usepackage{dsfont}
\usepackage{caption}
\usepackage{xcolor}
\usepackage{mathrsfs}
\usepackage{bbold}

\usepackage{mathtools}
\usepackage{graphicx,psfrag,subfigure}

\newcommand{\BEQ}{\begin{equation}}
\newcommand{\EEQ}{\end{equation}}
\newcommand{\bea}{\begin{eqnarray}}
\newcommand{\eea}{\end{eqnarray}}
\newcommand{\be}{\begin{equation}}
\newcommand{\ee}{\end{equation}}
\newcommand{\BEA}{\begin{eqnarray}}
\newcommand{\EEA}{\end{eqnarray}}

\newcommand{\nn}{\nonumber }

\begin{document}

\title{Semiclassical origin of suppressed quantum chaos in Rydberg chains}

\author{Markus M{\"u}ller}
\email{Markus.Mueller@psi.ch}
\affiliation{PSI Center for Scientific Computing, Theory and Data, CH-5232 Villigen PSI, Switzerland}
\author{Ruslan Mushkaev}
\affiliation{Department of Physics, University of Fribourg, 1700 Fribourg, Switzerland}

\date{\today}

\begin{abstract}
The surprisingly long-lasting oscillations observed in the dynamics of highly excited states of chains of Rydberg atoms defy the expectation that interacting systems should thermalize fast. The phenomenon is reminiscent of wavepackets in quantum billiards that trace classical periodic orbits.  
While analogs of the associated scarred eigenfunctions have been found for Rydberg chains, an underlying classical limit hosting periodic orbits has remained elusive. Here we generalize the Rydberg (pseudospin $S=1/2$) system to a chain of arbitrary spin $S$. Its classical limit features unexpectedly stable periodic orbits that are essential to understand the emergence of robust, parametrically suppressed quantum chaoticity, with semi-classical coherence times diverging as $\sqrt{S}$. The classical limit successfully explains several empirical features of the quantum limit.     
\end{abstract}

\maketitle

\section{Introduction - }
Generic many-body systems typically relax to local equilibrium within a few collision times.~\cite{thermalization} While a lot of recent research has focused on strongly coupled systems where this equilibration is particularly fast~\cite{fastscramblers1}, saturating a bound on the emergence of chaos~\cite{fastscramblers2}, it is equally important to understand when and why equilibration is particularly slow or even remains incomplete for all times. This is of particular interest when one aims to retain information of initial states, avoiding its loss due to local entropy maximization. 
Standard routes to preserving information for long times in many-body systems encompass the spontaneous symmetry breaking underlying conventional memories, including glassy systems and neural networks~\cite{glasses}, many-body localization in strongly disordered systems~\cite{mbl1,mbl2} or integrable systems~\cite{integrability}, which conserve extensively many quantities, as well as constrained systems with fragmented Hilbert spaces~\cite{fractons}.

Recently, a new subtle phenomenon that challenges the ergodic hypothesis, was discovered in a pioneering experiment on Rydberg chains where quasiperiodic motion starting from particular initial states persists for unexpectedly long times.~\cite{Lukin} 
Subsequently, many models have been identified~\cite{spin1XY,Pakrouski,AKLT1,AKLT2,Hubbard}, that feature non-decaying, exactly periodic dynamics if initialized within a specific subspace~\cite{ShiraishiMori,ReviewScars}. In most of these solvable cases the eigenstates participating in the dynamics were found to be highly atypical, defying the eigenstate thermalization hypothesis~\cite{ETH1,ETH2} (ETH) which postulates that quantum chaos induces a mixing of all thermodynamically relevant configurations such that local density matrices of individual eigenstates look like the restriction of the equilibrium Gibbs ensemble. Due to its similarity with the dynamics and eigenstates of quantum billiards that exhibit traces (so-called scars) of classical periodic orbits~\cite{Heller,Bogomolnyi88}, the many-body phenomenology was dubbed "quantum many-body scarring". Weakly unstable periodic orbits of billiards indeed ensure the slow spreading of wavepackets prepared close to them, which in turn implies that some eigenfunctions concentrate on the classical periodic orbits, their phase space density deviating from the uniform density naively expected from quantum chaos. 
However, rather than captalizing on this similarity and identifying a  suitable classical limit, most theoretical approaches to the many-body problem have so far focussed on constructing special, non-thermalizing "scar" subspaces, that are annihilated by most of the terms of the Hamiltonian, such that the dynamics within those subspaces become nearly trivial~\cite{ShiraishiMori,ReviewScars}. Similar structures were sought in the Rydberg problem and modified versions thereof~\cite{Turner, TurnerZ4, LinMotrunich18, Choi, Omiya1,Omiya2}. Since this algebraic approach remained rather far from the physics underlying low-dimensional quantum billiards~\cite{Heller}, the question arose, whether the two "scar" phenomena are genuinely related. 
Several attempts have recently been made to approximately project the many-body dynamics to low-dimensional configuration spaces, that could then be analyzed in terms of regular or chaotic dynamics~\cite{Ho, Pichler22, Richter23,Evrard24}. 
A recent study~\cite{Knolle24} retained the high dimensionality of the many-body problem and analyzed the classical large spin limit of quantum spin chains.
Those were found to host numerous unstable periodic orbits which generically show in quantum eigenstates by a slightly enhanced weight close to the classical orbits - as in billiards -  albeit with exponentially tiny amplitude in the thermodynamic limit~\footnote{It is not clear how crystalline symmetries of the initial state could protect against this exponentially weak scarring, as was suggested in Ref.~\cite{Richter23}.}, still compatible with the ETH. However, no direct consequences regarding slow thermalization were reported. 

Notwithstanding these insights, 
it has remained an open question whether the Rydberg chain and other slowly thermalizing models can be seen as emerging from a semi-classical limit with periodic orbits. Here, we close this gap by revealing an essential, but hitherto missed mechanism entailing parametrically suppressed quantum chaoticity. The latter will be shown to ensue from classical parent orbits that, unlike Heller's billiard orbits, enjoy an extraordinary stability. At the same time they seem rather oblivious to eigenstate scarring, which has indeed been argued to disappear under perturbations, while slow thermalization dynamics is the actual fundamental and robust phenomenon that needs to be explained.~\cite{LinChandran20,AsymptoticQMBS}

\section{Semiclassical limit}
We introduce an extension of the quantum Rydberg chain that admits a natural classical limit retaining a thermodynamically large number of degrees of freedom. This limit will be shown to host surprisingly many periodic orbits, which have the potential to cause slow thermalization in the quantum limit.

In the Rydberg chain of Ref.~\cite{Lukin} the atoms undergo laser-driven Rabi oscillations (single atom Rabi frequency $J$)  between a ground (0) and an excited state (1), provided their neighbors are unexcited, since otherwise their van der Waals interaction detune the transition from resonance. 
Associating states 0,1 with the levels $s^z=(-1/2,+1/2)$ of an effective $S=1/2$ spin, the Hamiltonian in the rotating frame of the driving laser reads
\bea
\label{HPXP}
H = 2J \sum_{i=1}^N P_{i-1}  s_i^x P_{i+1} 
\eea
where the projector
\bea
P_i = S- s^z_{i}
\eea
forces atom $i$ to be in the ground state ($s_z= -S\equiv -1/2$). We choose the energy units such that $2J=1$. 
Upon initializing the system in certain density waves (the $Z_2$ configuration (0101...), or a $Z_3$ state (001001...)) quasi-periodic oscillations were observed before the system eventually equilibrated. No oscillations were, however, observed for $Z_{n\geq 4}$ initial conditions, a fact that has so far remained unexplained. For convenience we consider periodic boundary conditions, identifying $s_{i+N}$ with $s_i$.

The model as written in Eq.~(\ref{HPXP}) immediately generalizes to spins of arbitrary size $S$~\footnote{As always, the generalization to larger $S$ and the $S\to \infty$ limit are not unique. Our analytic choice for generalizing the projector  is more amenable to a smooth large $S$ limit than the singular form $P= \delta_{s^z,-S}$ used in Ref.~\cite{Turner}.}. 
As usual the limit $S\to \infty$  defines a classical system whose quantum fluctuations are suppressed as $1/\sqrt{S}$. While a similar limit  was recently considered for chains with pairwise interactions~\cite{Knolle24}, the Rydberg system behaves fundamentally differently, leading us to very different conclusions about the relevance of the classical limit.  

The quantum dynamics is governed by the Heisenberg equations of motion, describing the precession of the spins in their local field 
\bea
\label{EQmotion}
\frac{d{\mathbf s}_i}{dt} = i[H(t),{\mathbf s}_i(t)] =  {\mathbf h}_i(t) \wedge {\mathbf s}_i(t),
\eea
with the components of the local field 
$ h_i^\alpha = -\frac{\partial H}{\partial s_i^\alpha}$.
These hold for any size of the spin $S$. In the limit $S\to \infty$, 
the Eqs.~(\ref{EQmotion}) become classical equations of motion for the expectation values ${\mathbf S}_i \equiv \langle {\mathbf s}_i \rangle/S$, which become unit vectors, while ${\mathbf h}_i$ turns into a classical field which depends on the nearest and next nearest neighbor spins. 

An important class of initial conditions are $n$-site periodic product states such as $|Z_n\rangle$. We therefore study the classical large $S$ dynamics initialized in such states. As their spatial periodicity is conserved by the time-evolution it suffices to describe the motion of the $n$ inequivalent spins within a unit cell. The motion involves $2n$ angular degrees of freedom, precessing under the first order equation of motion (\ref{EQmotion}). Comparing with two coupled penduli having 4 degrees of freedom, one might expect chaotic dynamics to ensue already for two-site periodic configurations ($n=2$). 
However, interestingly this analogy does not apply to the full phase space. For dynamics at total energy $E=0$ (the energy of the $Z_n$ states, in the center of the many-body spectrum),  
the dynamics are constrained and stabilized by a kind of particle-hole-like symmetry under a global rotation by $\pi$ around the $z$-axis, $R_z = \prod_i \exp(-i \frac{\pi}{2}  s_i^z)$,
acting on the quantum Hamiltonian as
\bea
\label{PHsymmetry}
R_z^\dagger HR_z = -H,
\eea
an operation that is used in experiments to implement time reversal and Hahn echoes~\cite{Maskara21}. This symmetry protects the dynamics of certain periodic states from developping classical chaos. 
Let us first analyze the case $n=2$, describing the motion of even and odd spins, resp.,
\bea
{\mathbf S}_{2m}(t)= {\mathbf S}_{e}(t); \\
{\mathbf S}_{2m+1}(t)= {\mathbf S}_{o}(t), 
\eea
whose motion is a priori not related in any simple way.
However, consider a spin configuration $\left\{ {\mathbf S}_{i} \right\}$ that maps onto itself upon acting by the generalized particle-hole symmetry ${\mathcal T} R_z$ where a spin rotation is combined with a translation by one site, ${\mathcal T}$, 
\bea
\label{SigmaDef}
({\mathcal T} R_z) \left\{ {\mathbf S}_{i} \right\} ({\mathcal T} R_z)^{-1}= \left\{ {\mathbf S}_{i} \right\}.
\eea
In those configurations ${\mathbf S}_{e}$ and ${\mathbf S}_{o}$ are thus related by a $R_z$-rotation.
The particle-hole symmetry \eqref{PHsymmetry} implies that their energy is $\langle H\rangle = E=0$. The set of all such configurations with period $n=2$ form a 2-dimensional manifold $\Sigma$.
If the system reaches $\Sigma$ at time $t_0$, the symmetry ensures that forward time evolution is identical to backward evolution, up to conjugation with ${\mathcal T} R_z$~\footnote{A similar observation also applies to the particle-hole-like symmetry $R_z$ alone. However, it only leaves 4 discrete configurations invariant, in which both $ {\mathbf S}_{e}$ and ${\mathbf S}_{o}$ are fully polarized along $z$.
}
\bea
\label{spinecho}
\left\{ {\mathbf S}_{i}(t_0+t)\right\} = 
{\mathcal T} R_z \left\{ {\mathbf S}_{i}(t_0-t) \right\} ({\mathcal T} R_z)^{-1}.
\eea
Upon reaching $\Sigma$ the system thus behaves similarly  as if a $\pi$-pulse had been applied in a Hahn echo, even though the refocussing takes place entirely due to the intrinsic dynamics.

\begin{figure}
    \centering
    \includegraphics[width=\linewidth]{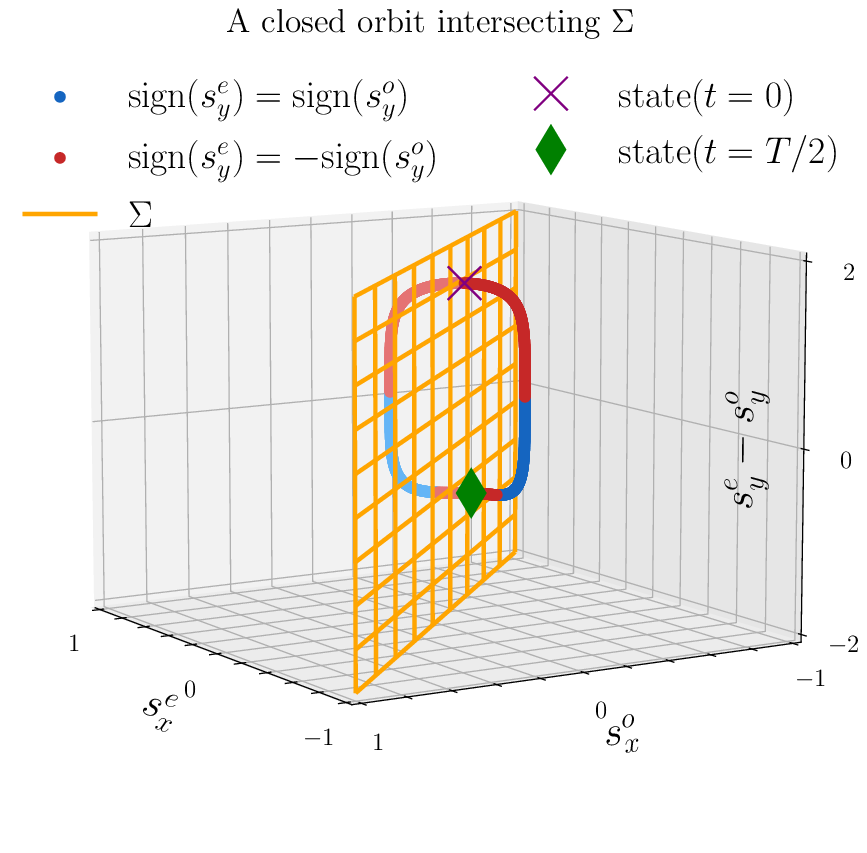}
    \caption{Two-site periodic trajectories crossing the echo manifold $\Sigma$ (orange plane). 
    {Forward time evolution from a point X on $\Sigma$ is equivalent to backward evolution followed by a one-site translation and a $\pi$-rotation about the $z$-axis, as reflected by the orbit's mirror symmetry with respect to the $\Sigma$-plane in the chosen parameterization ($s_{x,y,z}^{o,e}$ referring to the $x,y,z$-components of the even and odd spins, resp.). Forward and backward trajectories thus close by reaching $\Sigma$ in the same point (marked by a green diamond). 
    }}
    \label{fig:manifold}
\end{figure}

\begin{figure}
    \centering
    \includegraphics[width=\linewidth]{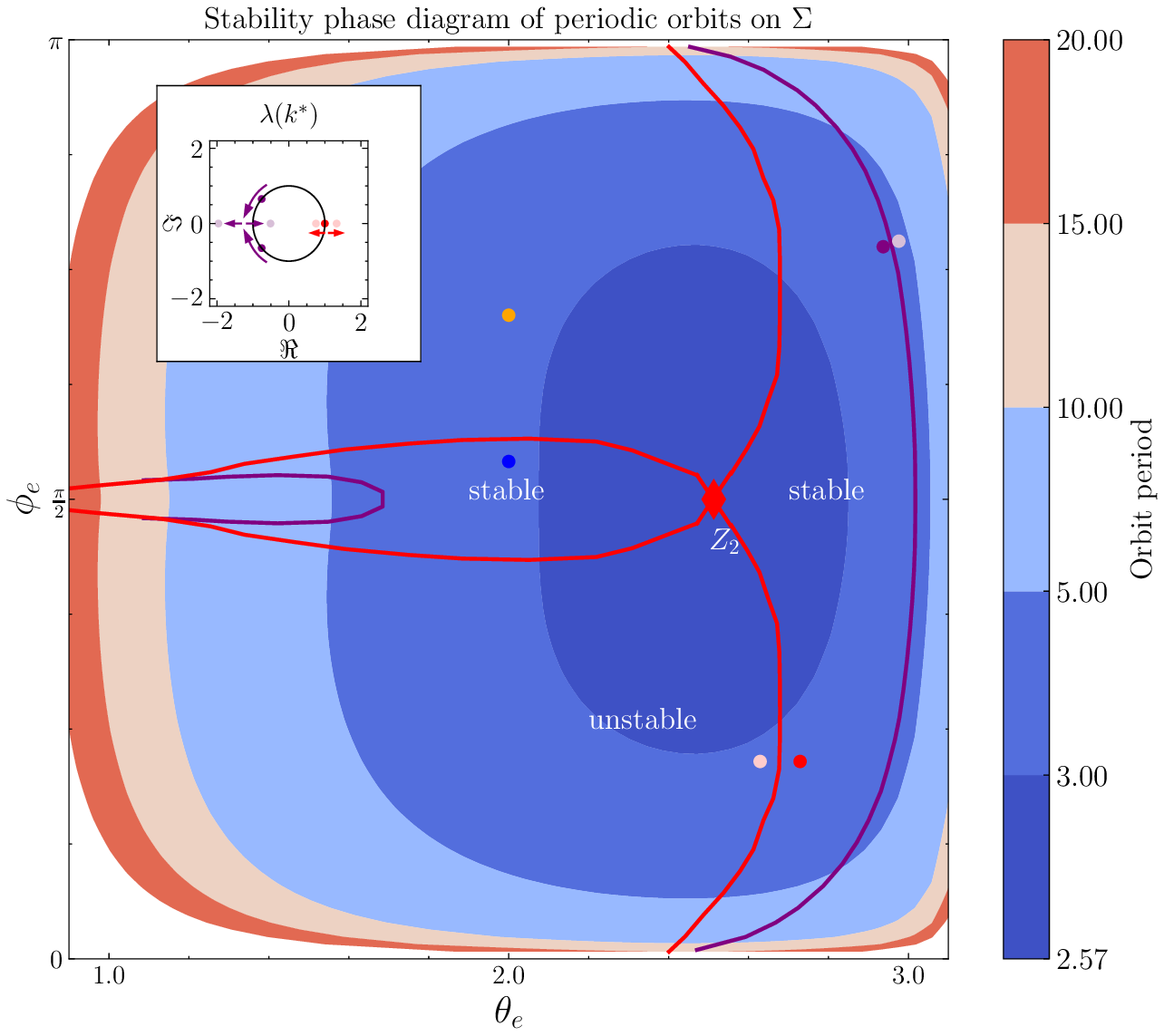}
    \caption{Stable and unstable regions of the two-site periodic orbits, labelled by the angles $\theta_e,\phi_e$ of even spins when the orbit crosses the echo manifold $\Sigma$. 
    The color encodes the orbit period (in units of $1/(2JS^2)$). The $Z_2$ orbit has the shortest period. Neighboring orbits with only small components $S^x_i$ ($\phi_e\approx \pi/2$) are fully stable. Red and purple boundaries indicate the two types of instabilities illustrated in the inset, where a pair of reciprocal eigenvalues $\lambda_{k^*},\lambda_{k^*}^{-1}$ of the stroboscopic forward map leaves the unit circle at $\lambda_{k^*}= \pm 1$, resp. The blue and orange points indicate  orbits whose dynamic growth and Lyapunov spectrum are shown in Figs.~\ref{fig:stable_unstable} and \ref{fig:tr_mk}.}
    \label{fig:stability}
\end{figure}

As the space of classical two-site periodic configurations of energy $E=0$ is 3-dimensional, the echo-manifold $\Sigma$ forms a hypersurface. Under time evolution it sweeps the entire $E=0$ configuration space, as we checked numerically. Moreover, the trajectory from any initial point $\{{\mathbf S}_{i}\} \in \Sigma$, will return to $\Sigma$ at a later time $t=T/2$, as illustrated in Fig.~\ref{fig:manifold}.
The echo property (\ref{spinecho})  implies that the orbit closes after a period $T$, since the points of the forward and backward return to $\Sigma$ coincide. 
The point where an orbit crosses the echo-manifold $\Sigma$ can be labelled by the polar angles ($\theta_e,\varphi_e$) of the even spins. They parametrize a two-parameter family of closed orbits. The orbit starting from the $Z_2$ configuration turns out to have the shortest period, corresponding to an angular frequency $\Omega_{Z_2}\approx 1.22 J \,(2S)^2$,  cf.~Fig.~\ref{fig:stability}. This is within 10\% from the experimental value $1.33 J$~\cite{Lukin} upon taking the quantum limit $S\to 1/2$.

An important subset of orbits is characterized by $S_i^x=0$ for all $i$, a condition  preserved under the dynamics.
The spins simply precess in the $y,z$-plane and are  characterized by the single angle $\theta_i$, obeying $ \dot{\theta}_i = (1-\cos \theta_{i-1})(1-\cos \theta_{i+1})$. 
This allows for special trajectories in a lower dimensional subspace: we find periodic orbits not only for $n=2$, but also for $n=3$, as ensured by additional conservation laws for generic configurations $\{\theta_i\}$. 
{From the equations of motion, it follows for $n=3$ that $\dot{\theta_i}[1-\cos(\theta_i)]-\dot{\theta_j}[1-\cos(\theta_j]=0$, implying the conservation laws $\theta_i -\sin(\theta_i)-\theta_j+\sin(\theta_j)={\rm const}$ $\forall i, j$. These constrain the dynamics to a 1-dimensional subspace, implying that all trajectories close.}  Trajectories  initialized in the $Z_2$ and $Z_3$ states belong to this family of closed orbits.
In contrast, the equations of motion for periodic configurations with $n\geq 4$ do not support any additional conservation laws. As a consequence the large number of angular degrees of freedom entails classical chaos. In particular the motion starting in  configurations $Z_{n\geq 4}$ does not close. The lack of a parent periodic orbit in the classical limit is in striking correspondence to the observed absence of long-lived quasiperiodic oscillations in the quantum limit.~\cite{TurnerZ4}

The mere presence of periodic orbits in the classical limit does, however, not yet imply any unusual dynamics, except for an exponentially weak scarring of eigenstates~\cite{Knolle24}. We rather need to analyze its semiclassical behavior. 
For single-particle billiards Heller~\cite{Heller} has shown that quantum eigenstates exhibit traces of classical periodic orbits only if the Lyapunov exponent $\lambda$ governing the exponential growth of small fluctuations from the periodic orbit is much smaller than the inverse of the orbit period. The analogous linear stability analysis for a  chain is more involved, as the dimensionality of the space of possible deviations grows with the system size. To characterize the stability of the periodic many-body orbits found above, we consider small fluctuations $\delta {\mathbf S}_i$ from a periodic configuration $\overline{{\mathbf S}}_i$ and 
forward propagate the spin configuration in time 
by the period $T$ of the orbit. The unperturbed periodic configuration is a fixed point of this stroboscopic map, $F_T$, while small flcutuations are propagated linearly under its tangential map $dF_T$, $\delta {\mathbf S}(T) = dF_T(\delta {\mathbf S}(0))$.  
In the spirit of a truncated Wigner approximation for the semiclassical limit at large $S$, typical quantum fluctuations away from the spin expectation value of unit norm scale as 
$\langle  \delta {\mathbf S}_{i}^2 \rangle(t=0) = 1/S.$

The dynamical growth of fluctuations is governed by the eigenvalues of $dF_T$ (the Lyapunov spectrum), 
which due to translational invariance acts diagonally in reciprocal space. 
Generically, one may expect a fraction of the eigenvalues to have modulus larger than 1, as  was reported for many simple spin chains~\cite{Knolle24}. This entails fast emergence of classical chaos, leaving little room for visible dynamical traces in the quantum limit. Interestingly, however, the Rydberg chain behaves very differently.




Indeed, as shown in Fig.~\ref{fig:stability}, a significant subset of the $n=2$ periodic orbits have forward maps whose eigenvalues are all pure phases, which prevents an exponential divergence away from the initial state. 
This is easy to see for the $Z_2$ orbit. Its tangential forward and backward maps are identical, since the $Z_2$ configuration is invariant under the rotation $R_z$, which inverses the sign of $H$ and thus the time evolution. 
(cf. also App.~\ref{app:stability}. 
Accordingly, one must have $dF_T^2= {\mathbf 1}$, implying that all  eigenvalues of $dF_T$ are $\pm 1$.   
A numerical analysis of general orbits shows that those with small spin components along $x$ tend to be stable, as well. 
Similarly as in billiards~\cite{KaplanHeller}, it has a stabilizing effect that the periodic orbits come in a continuous family. Here this assures that long wavelength fluctuations never grow rapidly.

The same arguments as for $Z_2$ also apply to the periodic orbit associated with the $Z_3$ state, which is equally anomalously stable as the $Z_2$ orbit. On the other hand, we recall that the $Z_4$ state does not have a closed periodic orbit, but instead leads to chaotic motion already within the manifold of $n=4$-periodic configurations. 

As we now show, the unusually stable orbits suppress the emergence of chaos in the semiclassical limit.



\section{Emergence of semi-classical chaos}

\begin{figure}
    \centering
    \includegraphics[width=\linewidth]{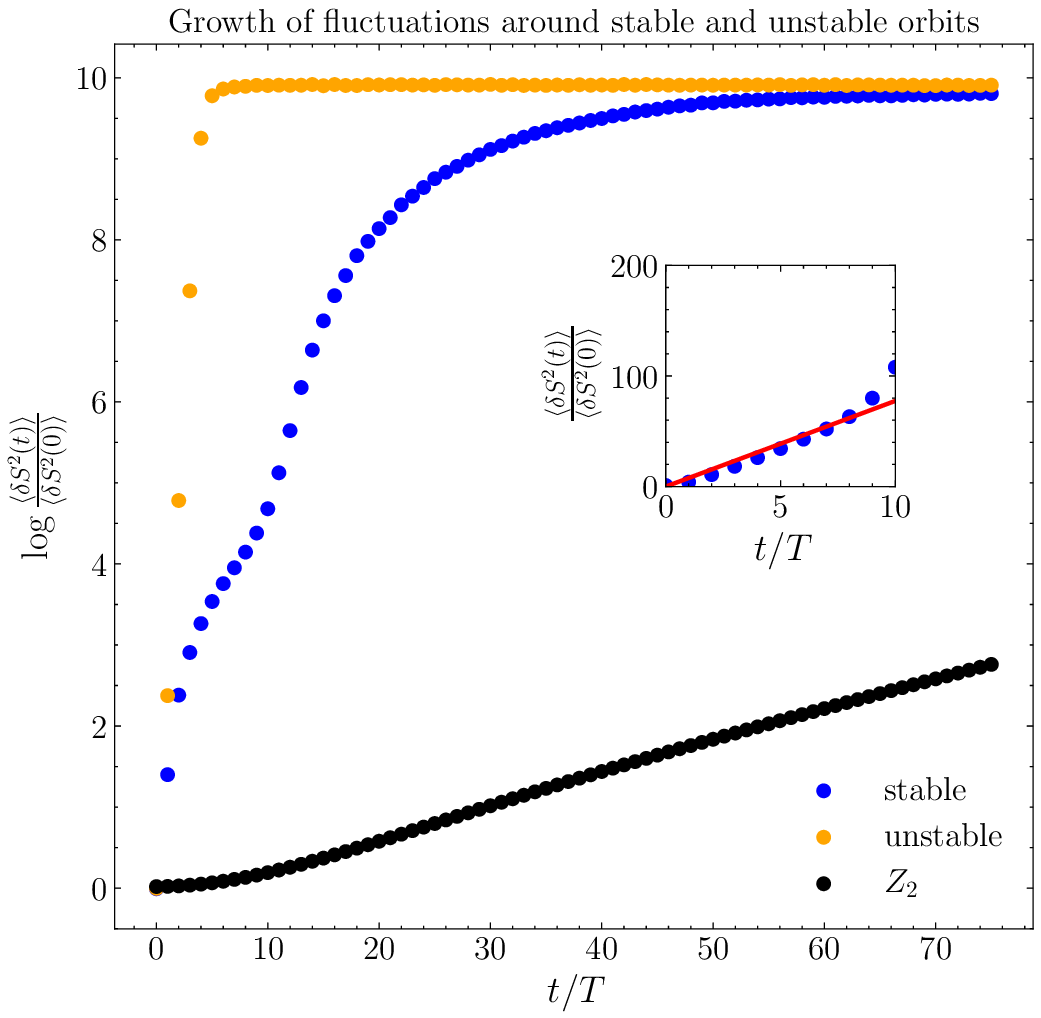}
    \caption{Growth of relative deviations from representative stable and unstable two-site periodic orbits, alongside the $Z_2$ state, all with initial fluctuations $\epsilon:= \sqrt{\langle \delta s_i^2(t=0) \rangle} = 0.01$. Unstable orbits exhibit exponential growth after $O(1)$ periods. Fluctuations of stable orbits generically grow linearly at small times (see inset), cf. Eq.\eqref{lineargrowth}, except the $Z_2$ orbit, and then turn to exponential growth with parametrically small growth rate.
    }
    \label{fig:stable_unstable}
\end{figure}

\subsection{Linearized time evolution}
{\em Unstable orbits - } If the stroboscopic map $dF_T$ has eigenvalues of modulus larger than unity, the fluctuations grow exponentially,  
\bea 
\langle \delta {\mathbf S}_i^2\rangle(t) \sim |\Lambda|^{t/T},
\eea
where $\Lambda$ is the largest eigenvalue of $dF_T$.
 This only leaves a time window $\sim\log(S)$, before initial spin fluctuations blow up and the motion turns entirely chaotic. Such unstable orbits are not expected to entail quasiperiodic oscillations in the quantum limit $S=O(1)$. 

\begin{figure}
    \centering
    \includegraphics[width=\linewidth]{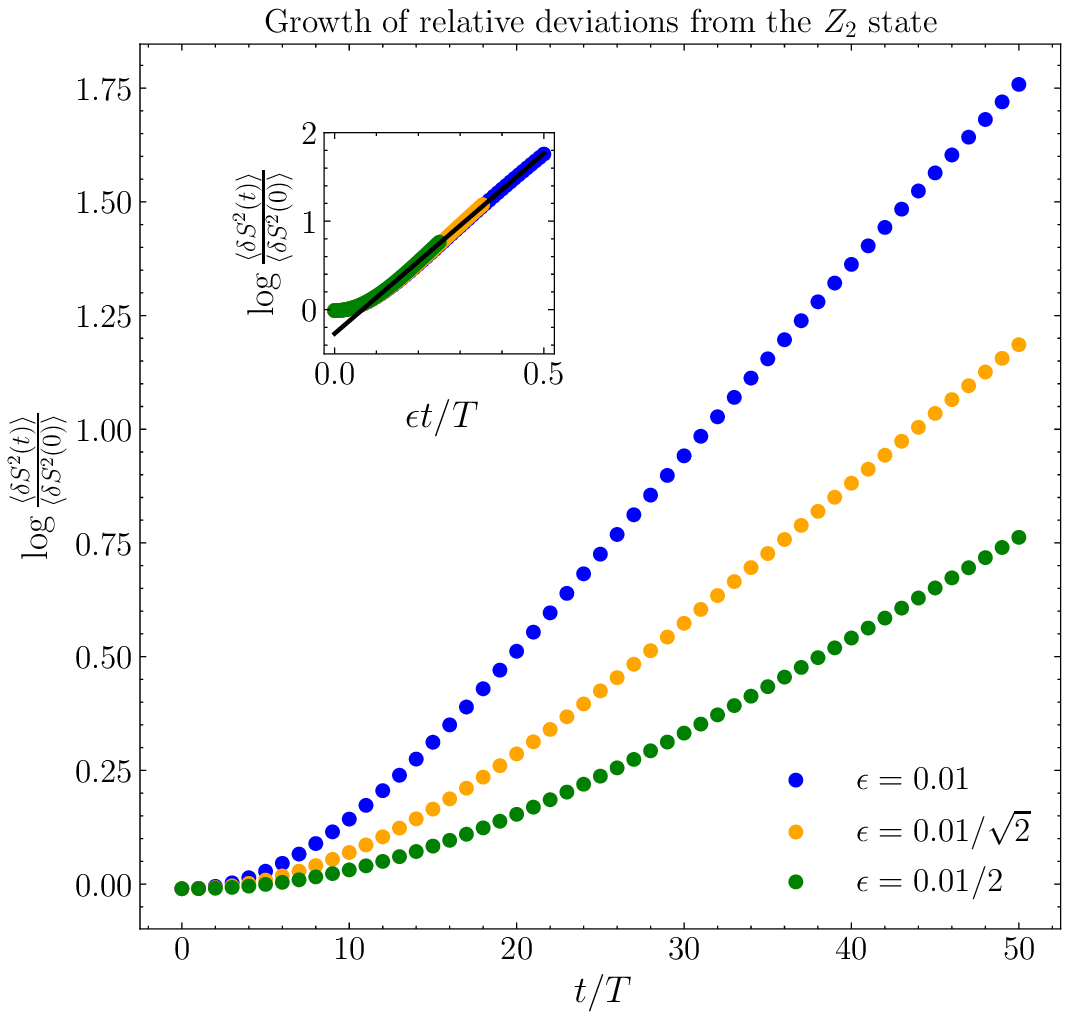}
    \caption{Exponential growth of random deviations from the classical $Z_2$ state for a chain of $N=100$ sites with initial fluctuations $\epsilon:= \sqrt{\langle \delta s_i^2(t=0) \rangle} = 0.01; 0.01/\sqrt{2}; 0.01/2$. After a crossover, the noise-averaged growth turns into an exponential (cf. inset), 
    with a parametrically suppressed growth rate $\propto \epsilon$. 
    }
    \label{fig:neel_ds}
\end{figure}

{\em Linearly stable orbits - } 
While the fluctuations around stable orbits do not grow exponentially, one still finds that the deviations from a local reference pattern grow algebraically with time. This is essentially due to the fact that generically the period of the considered orbit differs from those of its neighbors. Thus, long wavelength perturbations that mimic a slightly different periodic pattern will oscillate faster or slower. These regions fall out of lockstep with the rest of the chain, with a linearly increasing time lag or advance. The only exception is the $Z_2$ orbit whose period is a local minimum, such that its neighbors in orbit space have nearly equal frequencies, cf.~Fig.~\ref{fig:neel_ds}. This minimizes the growth of fluctuations for the $Z_2$ orbit.

A detailed analysis (cf. App.~\ref{app:stability}) shows that generically the fluctuations initially grow as $\delta S^2(t) \sim t$, as numerically confirmed in Fig.~\ref{fig:ds_scaling_stable}. 
However, as we now show, before the fluctuations grow to $O(1)$, non-linearities in the forward map kick in, which ultimately limits the semiclassical coherence time. 

\subsection{Chaoticity from non-linear effects}
For simplicity we focus on the $Z_2$ orbit, even though other linearly stable orbits can be treated similarly (cf. App.~\ref{app:exponential}).

At next-to-linear order, one  finds for the propagation of semiclassical fluctuations
\begin{equation}
    \delta {\bf S}(t+T)= \delta {\bf S}(t) + F_T^{(2)}[\delta {\bf S}(t)] + O(\delta {\bf S}^3),
\end{equation}
 where $F_T^{(2)}$ is the quadratic term in the Taylor expansion of the stroboscopic forward map. 
 
For the evolution of small fluctuations scaling as $||\delta {\bf S}||=\epsilon$, one can approximate the stroboscopic evolution by a continuous differential equation,
\begin{equation}
    \frac{d(\delta {\bf S})}{d(t/T)}\approx  F_T^{(2)}[\delta {\bf S}(t)] + O(\delta {\bf S}^3).
\end{equation}
Averaging over initial conditions with constant standard deviation 
{$\epsilon := \sqrt{\langle \delta {\bf S}^2(0)\rangle/N}$}, and rescaling $\delta {\bf S}$ and time by $\epsilon$, one finds that, as long as they remain small, the fluctuations must follow a scaling law
$\frac{||\delta {\bf S}(t)||^2}{||\delta {\bf S}(0)||^2} = \Phi(\epsilon \,t/T)$.
Numerical evaluation and averaging confirms this, showing moreover that after an initial crossover the scaling function $\Phi$ becomes a simple exponential, $\Phi(x \gtrsim 0.3) \approx \Phi_0 \exp(\kappa x)$ with $\Phi_0\approx 0.763$ and $\kappa \approx 4.07$. 

In App.~\ref{app:exponential} we argue that exponential growth is indeed a possible consistent growth pattern. If it indeed establishes, the above implies that the growth rate has to scale as $\epsilon$.
This can  be rationalized alternatively by noting that locally small deviations from the $Z_2$ orbit will often look like a configuration of a neighboring, slightly unstable orbit (or more generally of a slightly non-closing, but still weakly unstable trajectory). One can show that the maximal eigenvalue of the associated forward map scales linearly with that local displacement $||\Delta|| \sim\epsilon$ in orbit space (cf. App.~\ref{app:instabilityclosetoZ2}). Fluctuations within such local patches are thus expected to grow exponentially with a parametrically small rate $\sim \epsilon$. 
This translates into a parametrically large time $T_{\rm coh} \sim \epsilon^{-1} \sim S^{1/2}$ before semi-classical fluctuations grow to order $O(1)$ and establish full chaos. The linearly stable classical orbits may thus be expected to be at the root of long-lasting quasi-periodic motion, certainly in the semiclassical regime, but possibly all the way down to the quantum limit $S=O(1)$, as empirically seems to happen for the $Z_2$ and $Z_3$ orbits of  the Rydberg chain.

The suppressed quantum chaoticity we find here is in perfect agreement with the analytically derived early time growth of out-of-time correlators in a solvable bosonic toy model introduced by Omiya~\cite{Omiya3}. His model admits a classical limit in terms of the number $N\gg 1$ of boson species, which takes the role analogous to $\sqrt{S}=\epsilon^{-1} \gg 1$ here. 
 Omiya shows analytically that quantum chaos close to the linearly stable  periodic orbits develops exponentially, but with a parametrically small growth rate, laying the above heuristics on a firm ground.

\section{Discussion}
Our large $S$ approach is the first successful attempt at defining a parametrically controlled classical limit whose periodic orbits may rationalize the peculiar quasiperiodic motion observed in the quantum regime of the Rydberg chain. Note that crucially, the periodic orbits we find are linearly stable, not only weakly unstable - unlike the classical parent orbits in billiards~\cite{Heller} - in contrast to conjectures in the recent literature~\cite{Richter23,Evrard24,Knolle24}. 

The large $S$ limit is a genuine, parametrically controlled classical limit that allows to tune and  suppress quantum chaoticity by dialling $S$. Our limit retains the full complexity of spatially inhomogenous fluctuations that drive certain periodic parent orbits unstable
As such it is rather distinct from approaches that use long range interactions and the resulting large collective moments to stabilize robust collective dynamics in a effectively low-dimensional space.~\cite{Pichler22, Pappalardi23,Evrard24}.

One may wonder how the fact that all the periodic orbits have energy  in the center of the many-body spectrum, $E=0$, could be compatible with the empiric observation that the tower of exceptional, low entangled 'scar' states of the quantum Rydberg chain extends across the entire bandwidth.~\cite{Lukin} 
This is resolved by noting that in the large $S$ limit, the classical oscillation frequency associated with a periodic orbit scales as $\Omega \sim S^2$. 
The associated tower of  anticipated  $N$ scar states thus  extend over a range $\sim NS^2$. In the large $S$ limit this is merely a small fraction of the width of the many-body spectrum $N S^3$, and it is thus natural that  the periodic orbits have energy density shrinking to $E/N\sim 1/S\to 0$ in the classical limit. Those orbits can nonetheless root scar states participating in slow thermalizing dynamics at finite energy density in the regime $S=O(1)$.

As was noticed in Ref.~\cite{Knolle24} Heller’s arguments for the low-dimensional case, lose most of their strength in the many-body case. The large dimensionality of the configuration space in the thermodynamic limit seems to prevent one from concluding that long-lived periodic motion of wavepackets imply eigenstates that defy the ETH in a significant way.
From our classical limit, we are indeed not able to infer significant scarring of eigenfunctions in the intermediate $S$ regime. In contrast the suppressed emergence of quantum chaoticity remains a robust feature. From this perspective it appears that strong scarring of eigenfunctions as found in many solvable toy models might be rather a property of their fine-tuned Hamiltonians, similarly as their infinite coherence time. The same conclusion is supported by Omiya's model~\cite{Omiya3}, where suppressed chaoticity persists even if small perturbations destroy the non-thermal nature of the exact scar eigenstates.
This offers a fresh perspective on  recent observations that slow thermalization might well be more robust than the existence of ETH-defying scar states by themselves~\cite{AsymptoticQMBS}.

Several interesting open questions remain, especially concerning the precise relation between stable orbits in a certain classical limit and exact low entanglement 'scar' eigenfunctions of  fine-tuned Hamiltonians in the quantum limit.
Does a classical model with linearly stable orbits necessarily entail the possibility of defining quantum deformations with ETH-defying eigenstates? Is there an 'optimal' way of defining a large $S$ (or other classical) limit for a given Hamiltonian with pronounced eigenstate scarring, and does the latter guarantee a classical limit hosting associated stable orbits?
Further insight into these questions will not only deepen our understanding of large $S$ approaches and the general mechanisms governing quantum chaos, but may also allow us to design realistic Hamiltonians with enhanced robustness with regard to the quantum revivals and the associated configurational memory. 


\acknowledgements
We thank K. Omiya for numerous discussions and sharing his insights prior to publication.
We also thank Z. Papi{\'c}, S. Pappalardi and K. Richter for useful discussions.
This work was supported by the Swiss National Science Foundation via NCCR Marvel and SNSF Grants No. 200020-200558 (MM) and 200021-196966 (RM).




\appendix



\section{Stability analysis}
\label{app:stability}
The stability of periodic orbits can be analyzed for general $n$-periodic orbits.
Translational invariance ensures that the eigenvectors of the linearized stroboscobic map $dF_T$ are plane waves with a spatial variation 
\bea
\delta {\bf S}_{nj+l} &=& \delta {\bf S}_{l}(k) \exp(i k n j), \, \forall j, \,  1\leq l \leq n, 
\eea
with $-\pi/n <k \leq \pi/n$.  

Linearizing the equations of motion around a periodic orbit $\overline{S}$ 
and integrating them over the period period $T(\overline{S})$ yields  the stroboscopic map, a $2n\times 2n$ matrix $dF_{k}$ which maps the initial deviations $\delta S(t=0)$ onto the deviations after one period, $\delta S(T)= dF_{k} \delta S(0)$.


\subsection{Constraints on the stroboscopic forward map for $n=2$-site periodic orbits}
For orbits with space period $n=2$, stroboscopic propagation from the echo manifold $\Sigma$, $dF_T^\Sigma$, obeys an additional  constraint since forward propagation is conjugate to backward propagation under the map $C\equiv {\cal T}R$ that shifts the chain by one lattice spacing and rotates all spins by $\pi$ around the $z$-axis. This leaves $\Sigma$ invariant and induces a linear map on its tangent space.
The resulting constraint on $dF_T$ reads 
\bea
\label{Sigma}
(dF_T^\Sigma)^{-1}= C^{-1}(dF_T^\Sigma) C.
\eea
The symplectic property of spin dynamics (cf. App.~\ref{App:symplectic}) imposes a further constraint, namely 
\bea
\label{constraintforSigma}
(dF^\Sigma_{T,k})^{-1} = J^{-1}(dF^\Sigma_{T,k})^T J,
\eea
where $J$ is the linear map that rotates all tangent vectors by $\pi/2$ around the spin of the reference periodic orbit.
Equating the two expressions one finds
\bea
(dF^\Sigma_{T,k})^T =  J C_k^{-1}dF^\Sigma_{T,k} C_k J^{-1},
\eea
where $C_k$ is the restriction of $C$ to the sector of wavevector $k$. 

The last relation implies that if $\psi$ is a right eigenvector (REV) of $(dF^\Sigma_{T,k})^T$ (and thus a left eigenvector (LEV) of $dF^\Sigma_{T,k}$), then  $C_k J^{-1}\psi$ is a REV of $M^\Sigma_k$ with the same eigenvalue. Another REV with the same eigenvalue can be obtained by constructing the vector $\phi$ that is orthogonal to all LEVs of $dF^\Sigma_{T,k}$ except for $\psi$. As $\phi$ and $C_k J^{-1}\psi$ will in general be linearly independent it follows that generically the eigenvalues are at least two-fold degenerate.
Further, from Eq.~(\ref{constraintforSigma}) it follows that the spectrum of $dF^\Sigma_{T,k}$ is invariant under inversion ($\lambda\to \lambda^{-1}$), and since $dF^\Sigma_{T,k}$ is real (cf. App.~\ref{App:symplectic}), it is also invariant under complex conjugation. These properties will be confirmed explicitly below.

The four eigenvalues of $dF^\Sigma_{T,k}$ are thus pairwise degenerate. For the two different eigenvalues $\lambda_k^{(1,2)}$ there are only two possibilities: Either they are pure phases,  $\lambda_k^{(1,2)}= e^{\pm i\phi_k}$, or they are real and mutual inverses,  $\lambda_k^{(1)}= 1/ \lambda_k^{(2)}$.
A periodic orbit is linearly stable if for all $k$ the $\lambda_k^{(1,2)}$ are pure phases. which ensures that to linear order perturbations do not grow exponentially under forward propagation. 
Since $\lambda_k^{(1,2)}$ are continuous in  $k$, the only way for an orbit to become unstable is that for a critical wavenumber $k=k^* \in (0,\pi)$ the pair of eigenvalues tends to $\lambda_{k^*}^{(1)}=\lambda_{k^*}^{(2)}= \pm 1$, so that for $k>k^*$ the pair $\lambda_k^{(1)}= 1/ \lambda_k^{(2)}$ moves to the real axis with the maximal eigenvalue having modulus larger than 1. 

\begin{figure}
    \centering
    \includegraphics[width=\linewidth]{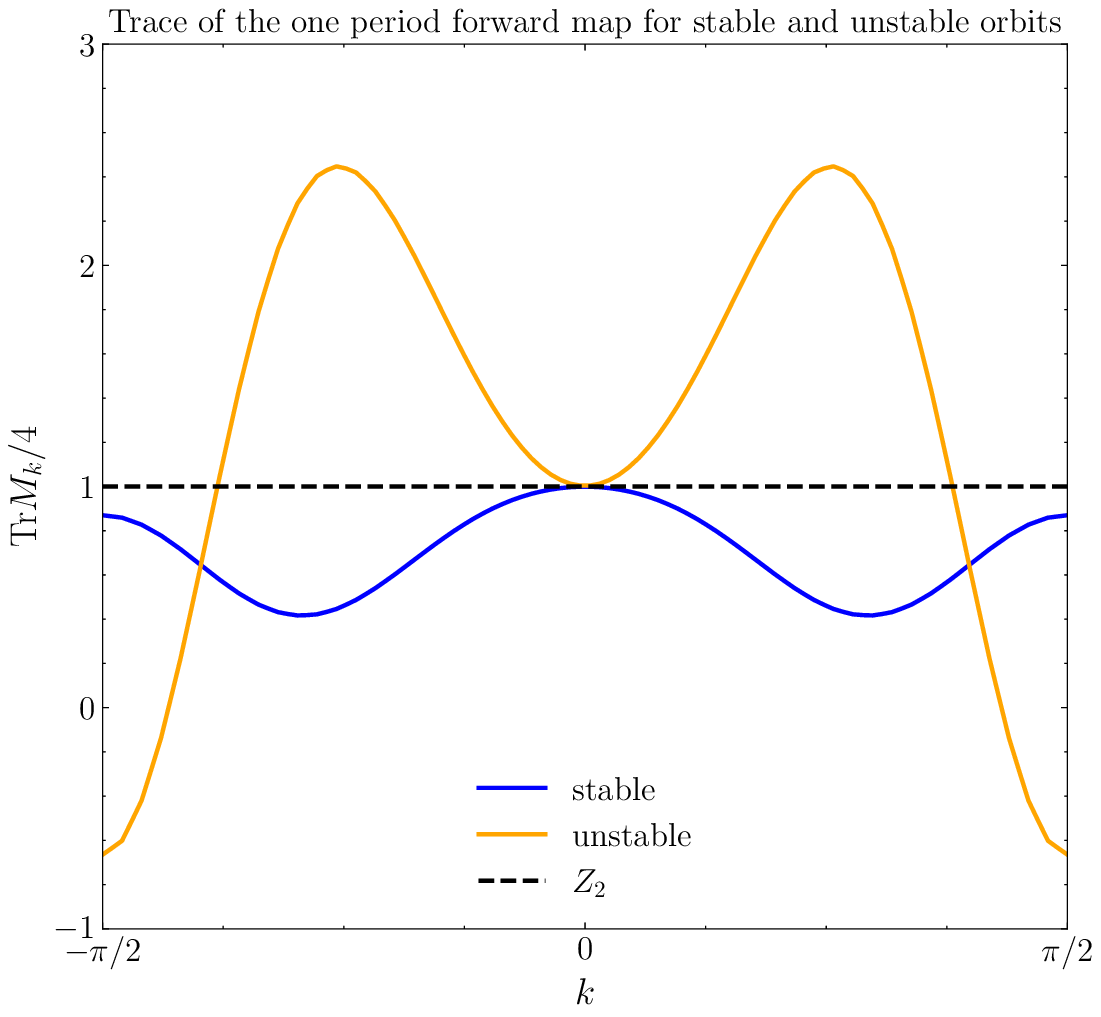}
    \caption{Trace of the one-period forward map $M_k$ as function of fluctuation wavevector $k$ for a typical stable and unstable orbit (indicated in Fig.~\ref{fig:stability} by the same color) of energy $E=0$. Stable orbits have $-1 \leq \mathrm{Tr}(M_k/4) \leq 1 \quad \forall k$. The $Z_2$ orbit is marginal as the eigenvalues of $M_k$ all equal 1 for any $k$. 
    }
    \label{fig:tr_mk}
\end{figure}

Fig.~\ref{fig:tr_mk} shows the traces of the stroboscopic map of fluctuations of wavevector $k$ $Tr(M_k)/4$ for a stable and an unstable orbit marked in Fig.~\ref{fig:stability}, as well as the $Z_2$ orbit. 
 Interestingly, the latter appears right at the border of orbit stability on $\Sigma$, which is manifested by the marginal situation $t_k=\lambda_k = 1\, \forall k$ (see main text).

\subsection{Structure of $dF_{T,k}^\Sigma$}
To express the stroboscopic map $dF_{T,k}^\Sigma$ as a matrix, we construct a convenient basis for the tangent space of a specific 2-site periodic configuration (c) on the echo manifold $\Sigma$. Consider translationally invariant ($k=0$) perturbations. 
Both the velocity vector $\tilde v^1 := \frac{dS}{dt}|_{t=0}$ and the vector in the tangent space of $c$ pointing toward another $E=0$ orbit with equal period, $\tilde v^3$, are $k=0$ eigenvectors of the forward map $M^\Sigma$ with eigenvalue 1. 
$\tilde v^1$ is odd under the time-reversing operation $C$, while $\tilde v^3$, as an infinitesimal distance between points that are invariant under $C$, is even and thus orthogonal to $\tilde v^1$. We complement the normalized vectors $v^{1,3} \equiv \tilde v^{1,3}/|| \tilde v^{1,3}||$ with vectors whose components on the site $i$ are rotated by $\pi/2$ around $s_i$, $v^2 := J v^3$ and $v^4 := J v^1$, to obtain a natural orthonormal basis of the sector $k=0$.~\footnote{We note that $v^4$ is proportional to the energy gradient, $v^4 \propto \nabla E$ evaluated within the manifold of fixed spin magnitude, $|s_i|=1$. Indeed, that gradient is odd under $C$ and  orthogonal to $v^1$ since the dynamics conserve the energy. Likewise,
$v^2$ is proportional to the gradient (taken within $\Sigma$) of the orbit period  (being even under $C$ and orthogonal to $v^2$, which points in the direction of constant period).}

From the above it follows that the component on the various sites are $v^1_j= [-R_z(\pi)]^j w$ and $v^3_j= [R_z(\pi)]^jz$, 
with two 3-vectors $w,z \perp s_0$, while $v^{(2,4)}_j= s_j\wedge v^{(3,1)}_j$. 
For perturbations at finite $k$ a convenient basis is $v^1_j(k)= ([-e^{-ik}R_z(\pi)]^j)w$ and $v^3_j(k)= [e^{-ik}R_z(\pi)]^j z$, $v^{(2,4)}= J v^{(3,1)}$ 
In the basis $v^{n=1,...,4}(k)$ the maps $C_k$ and $J$ take the simple forms (using Pauli matrices $\sigma^\alpha$)
\bea
J = \left(
\begin{array}{cc}0 & -i\sigma^y \\
-i\sigma^y & 0 
\end{array}
\right); \, 
C_k = e^{ik}\left(
\begin{array}{cc}-\sigma^z & 0 \\
0 & \sigma^z 
\end{array}
\right).
\eea

The constraints of Eqs.~\eqref{Sigma} and \eqref{constraintforSigma} impose the following structure on $dF_{T,k}^\Sigma$ for every wavevector $k$:
\bea
\label{Mk}
dF_{T,k}^\Sigma = \begin{pmatrix}
    A & b\sigma^z \\ c\sigma^z & A
\end{pmatrix}_k ; \, {\rm with }\,\, A = \begin{pmatrix}
    a & f \\ g & a
\end{pmatrix},
\eea
where $a,f_\pm,b,c$ are real, analytic and even functions of $k$, obeying 
$fg= a^2-1-bc$.
For $k=0$, the fact that $v^1$ and $v^3$ are eigenvectors of $dF_{T,k=0}^\Sigma$ imposes $b(k=0)=c(0)=g(0)=0$ and $a(0)=1$, so the matrix takes the form: 
\bea
\label{Mk0}
dF^\Sigma_{T,k=0} = \begin{pmatrix}
    1 &  f_0 & 0 & 0 \\ 0 & 1 & 0 & 0 \\ 0 & 0 & 1 &  f_0 \\ 0 & 0 & 0 & 1
\end{pmatrix},
\eea
where $f_0\equiv f(k=0)$ quantifies by how much the orbit period changes as one considers neighboring orbits (moving along $v^2$ on $\Sigma$).
The $Z_2$ orbit is special as it has the minimum orbit period and thus $f_0=0, dF^\Sigma_{T,k=0}= \mathbb{1}$ (consistent with the result derived in the main text). For all other orbits one finds $f_0\neq 0$, and the stroboscopic forward map in the sector $k=0$ is thus non-diagonalizable. As we will see this affects the initial growth of fluctuations around the ideal 2-site periodic orbits characterized by the condition that $|a(k)|= |{\rm Tr}(dF_{T,k}^\Sigma)|/4\leq 1 \,\forall k$.





\subsection{Linearized growth of fluctuations around linearly stable orbits}

Stable orbits intersecting $\Sigma$ are chracterized by pure phases as eigenvalues of $dF_T^\Sigma$. Naively one might thus expect fluctuations not to grow at all in time. However, even though exponential growth is indeed absent, the deviations from the periodic reference configuration grow linearly with time, as a consequence of the stroboscopic map $M^\Sigma_{k=0}$ being non-diagonalizable. 
To see this, let us consider a perturbation of a stable orbit on $\Sigma$ by a Gaussian-distributed $\delta s_i= S_i-\overline{S_i}$ of variance $\langle \delta s_i^2 \rangle = \epsilon^2=1/S \ll 1$ and zero mean $\langle \delta s_i \rangle=0$ as expected for quantum fluctuations of spins of size $S$. In the spirit of a truncated Wigner approximation a quantum wavefunction can be seen as a superposition of such fluctuating configurations.

We are interested in the growth of the initial fluctuations after $n$ periods $T$ of the underlying orbit, that is $\frac{\langle \delta S^2(n T)\rangle}{\langle \delta S^2(0)\rangle}$, where $\delta S^2 := \sum_j \delta s_j^2$ 
and $\langle . \rangle$ denotes an average over Gaussian statistics. 
We decompose the initial fluctuations into the eigenmodes $u_k^\alpha$ of the stroboscopic map $dF_T^\Sigma$, writing 
$\delta s_j = \sum_{k} \sum_{\alpha=1}^4 \delta_{\alpha,k} u^\alpha_{k,j}$.

The time-evolved fluctuation  $\delta S^2(n T)$ then evaluates to:
\bea
\label{ds2}
\delta S^2(n T) = 
\sum_{k}\sum_{\alpha,\beta=1}^4  u_k^{\alpha*} \cdot  u^\beta_k \langle \delta_{\alpha,k}^{*} \delta_{\beta,k} \rangle (\lambda_\alpha^{k*} \lambda_\beta^k)^n. 
\eea

Using the form of $dF_k$ derived in Eq.~\eqref{Mk} and diagonalizing it for small $k$ one obtains the nearly degenerate, non-orthogonal eigenbasis: $u_k^\alpha = \{(1,\pm i \mu k,0,0),(0,0,1,\pm i \mu k)\} +O(k^2)$, both associated with the eigenvalues $e^{\pm i \phi k}$, with  coefficients $\mu = \sqrt{\frac{- g''(0)
}{2f_0}}, \phi = \sqrt{-f_0 g''(0)/2}$. \footnote{$\mu$ and $\phi$  are real since $f_0 g''(0)= 2a''(0)<0$ for stable orbits for which $a(k)\leq 1=a(0)$.} Expressing the expansion coefficients $\delta_{\alpha,k}$ in terms of the independent Gaussian components of $\delta s_j$ and averaging over those, one obtains for the growth of deviations


\bea
\frac{\langle \delta S^2(t=n T)\rangle}{\langle \delta S^2(0)\rangle} 
& \approx & 1+\int_{-\pi}^\pi \frac{dk}{2\pi}\sin^2(\phi k n) \left[\frac{1}{(\mu k)^2}+O(k^0)\right] \nonumber\\
&\stackrel{n\gg 1}{\simeq} & \frac{1}{2} \frac{|\phi|}{\mu^2}n = \frac{ f_0^{2}}{[-2f_0g''(0)]^{1/2}}n.
\label{lineargrowth}
\eea
For small deviations in the linear regime, fluctuations thus grow linearly in time $t= nT$, except for the $Z_2$ orbit for which $f_0=0$. This growth reflects long wavelength perturbations that result in locally altered periodic configurations having a different period from the rest of the chain.


\section{Symplectic property of spin dynamics}
\label{App:symplectic}
In a similar way as the monodromy matrix  in classical dynamics is a symplectic matrix, classical spin dynamics underlies the following constraint:
\bea
\label{constancy}
\sum_i ({\mathbf S}_i \wedge \delta {\mathbf S}^{1}_i)\cdot \delta {\mathbf S}^{2}_i={\rm const.}
\eea 
Here, $\delta {\mathbf S}^{1,2}_i$ are two linearly independent infinitesimal perturbations belonging to the tangent space of the configuration. 
To show the above we start from the equation of motion
\bea
\frac{d}{dt}{\mathbf S}_i= {\mathbf h}_i \wedge {\mathbf S}_i
\eea
with ${\mathbf h}_i^\beta = \frac{\partial H}{\partial {\mathbf S}_i^\beta}$. Therefore,
the infinitesimal perturbations satisfy equations that follow from a linearization of the full equation of motion, i.e., 
\bea
\frac{d}{dt}\delta{\mathbf S}^\alpha_i &=& \sum_{\beta,\gamma}\epsilon^{\alpha \beta \gamma}\sum_{j,\nu} \left[\left(\frac{\partial {\mathbf h}^\beta_i}{\partial  {\mathbf S}_j^\nu} \delta{\mathbf S}_j^\nu\right)  {\mathbf S}_i^\gamma+{\mathbf h}_i^\beta  \delta{\mathbf S}_i^\gamma\right]\nonumber\\
&=&  \sum_{\beta,\gamma,j,\nu}\epsilon^{\alpha \beta \gamma} \left[\left(\frac{\partial^2 H}{\partial  {\mathbf S}^\beta_i \partial {\mathbf S}_j^\nu} \delta{\mathbf S}_j^\nu\right) {\mathbf S}^\gamma_i
+{\mathbf h}_i^\beta  \delta{\mathbf S}_i^\gamma \right],\nn
\eea
where $\epsilon_{abc}$ is the totally antisymmetric tensor. Using this in the time derivative of the RHS of Eq.~(\ref{constancy}), as well as the symmetry of the second derivative of $\frac{\partial^2 H}{\partial  {\mathbf S}_i^\beta \partial {\mathbf S}_j^\nu}$ under exchange $(i,\alpha) \leftrightarrow (j,\beta)$ and the orthogonality of ${\mathbf S}_i$ and the tangent vectors $\delta {\mathbf S}_i^{1,2}$, one confirms that the LHS of Eq.~(\ref{constancy}) is indeed constant in time.

Consider the linear map $\hat{J}$ which rotates deviations $ \delta {\mathbf S}_i$ around ${\mathbf S}_i$,
$(\hat{J} \delta {\mathbf S})_i= {\mathbf S}_i\wedge \delta {\mathbf S}_i$.
For the stroboscopic forward map $dF_T$ along a periodic orbit, the above constraint implies $dF_T^T \hat{J} dF_T=\hat{J}$. 

 Due to translational invariance $dF_T$ is block diagonal in reciprocal space, the restriction $M_k$ being a $2n \times 2n$ matrix for an $n$-site periodic configuration. The symplectic constraint implies 
 \bea
 M_k^{*T} J M_k = J,
 \eea
 where $J$ is the action of $\hat J$ restricted to an $n$-site unit cell. 
 
 For an orbit that is symmetric under inversion of site indices $i\to -i$ (as, e.g., for $n=2$), one has $M_k=M_{-k}= M_{k}^*$, from which it follows that $M_k$ is real.

\section{Nonlinear evolution of fluctuations under stroboscopic map - parametrically slow exponential growth}
\label{app:exponential}
To understand the semiclassical dynamics beyond linear stability we Taylor expand the stroboscopic forward map to second order,
\bea
\delta s_{k,\alpha}(t+T)&=& \lambda_{k,\alpha} \delta s_{k,\alpha}(t)\\
&&+\sum_{k',\beta, \gamma} F_{k,k',k''}^{\alpha,\beta \gamma} \delta s_{k',\beta}\delta s_{k''=k-k',\gamma},\nn
\eea
where $\delta s_{k,\alpha}$ is the amplitude of the fluctuation eigenmode with wavevector $k$ and mode label $1\leq \alpha \leq 4$.  
We assume $\delta s\ll 1$. We take the linear evolution into account by setting $\delta \hat s _{k,\alpha}(t):= \lambda_{k,\alpha}^{-t/T} \delta s_{k,\alpha}(t)$, so that the forward propagation only contains small quadratic terms,  justifying a continuum limit,
\bea
\dot{\delta \hat s}_{k,\alpha} = \lambda_{k,\alpha}^{-1} \sum_{k',\beta, \gamma} \left(\frac{\lambda_{k',\beta}\lambda_{k'',\gamma}}{\lambda_{k,\alpha}}\right)^\tau F_{k,k',k''}^{\alpha,\beta \gamma} \delta \hat s_{k',\beta}\delta \hat s_{k'',\gamma},
\eea 
where $\tau= t/T$ and $\dot \,\equiv  d/ d\tau$. For a generic stable orbit, the $\lambda_{k,\alpha}$ are pure phases, that depend on $k$. For $\delta \hat s\ll 1$  $\tau$ becomes large, before significant dynamics sets in. Accordingly, the sum over $k'$ can be approximated by a saddle point method, selecting the values of $k'$, for which there are indices $\beta$ and $\gamma$, such that  
$\frac{\lambda_{k',\beta}\lambda_{k-k',\gamma}}{\lambda_{k,\alpha}}=1$. Generically, there will be $O(1)$ solutions satisfying this constraint.
For simplicity we assume that there is just one solution $k'$ for a given $k$. To further simplify, we assume that there are only two modes $\alpha=\pm$ (necessarily implying that their associated phases $\lambda_k^- = (\lambda_k^+)^*=1/\lambda_k^+$ are complex conjugates), and we choose the labels $\alpha$ such that $(k,+)$, $(k',+)$ and $k'',+$ form a 'resonant' triple ($k''\equiv k-k'$).
The equation for ${\delta \hat s}_{k',+}$ receives a saddle point contribution from the pair of modes $[(k,+),(-k'',-)]$ (which we denote by $(k+, k''-)$), etc. 

Within the above approximations one then finds the set of 6 coupled non-linear ODEs  
\bea
\dot{\delta \hat s}_{k+} &=& A_{k+} \delta \hat s_{k'+}\delta \hat s_{k''+},\\
\dot{\delta \hat s}_{k'+} &=& A_{k'+} \delta \hat s_{k+}\delta \hat s_{k''-},\\
\dot{\delta \hat s}_{k''+} &=& A_{k''+} \delta \hat s_{k,+}\delta \hat s_{k'-},\\
\dot{\delta \hat s}_{k-} &=& A_{k-} \delta \hat s_{k'-}\delta \hat s_{k''-},\\
\dot{\delta \hat s}_{k'-} &=& A_{k'-} \delta \hat s_{k-}\delta \hat s_{k''+},\\
\dot{\delta \hat s}_{k''-} &=& A_{k''-} \delta \hat s_{k-}\delta \hat s_{k'+},
\eea
where we used that the modes at $k$ and $-k$ share the same phases, $\lambda_{k,+}=\lambda_{-k,+}$, and the $A_{k\pm}$ are amplitudes resulting from the saddle point approximation.

Note that if $\delta \hat s(\tau)= F(\tau)$ solves the initial value problem for  given $\delta \hat s(0)$, then 
the solution of the problem with rescaled initial  values $\delta \hat s'(0)=  \epsilon \hat s(0) $ is given by 
$\delta \hat s'(\tau)= \epsilon F(\epsilon \tau )$. 
Asymptotically one expects some components of $F$ to grow at least exponentially with time. The coherence time should be defined by some components of $\delta \hat s$  growing to $O(1)$, which  order of magnitude wise occurs when $F$ turns to its exponential growth. The rescaling argument then shows that the mean coherence time scales inversely with the size of the initial conditions, which in our semiclassical approach are given by $T_{\rm dec}\sim 1/\delta \hat s_{\rm typ} \sim {\sqrt S}$ which is parametrically large. 

Empirically, we find that the solution of the full non-linear forward propagation indeed turns eventually to a simple exponential growth. An exponential growth pattern indeed appears to be consistent  if we suppose that $ \delta \hat s_{k,\pm}(\tau)\sim c_{k,\pm}\exp(\pm\omega_k \tau)$,
with a triple of rates $\omega_{k,k',k''}$ that satisfy $\omega_k=\omega_{k'}+\omega_{k''}$. 
This ansatz does not fix the exponential growth rate, however. Indeed, as the previous consideration shows, rescaled initial conditions lead to a rescaled growth rate. It thus remains an open problem to analytically predict the value of the ratio $T_{\rm dec}/\sqrt{S}$, which will depend on the orbit considered. 

\begin{figure}
    \centering
    \includegraphics[width=\linewidth]{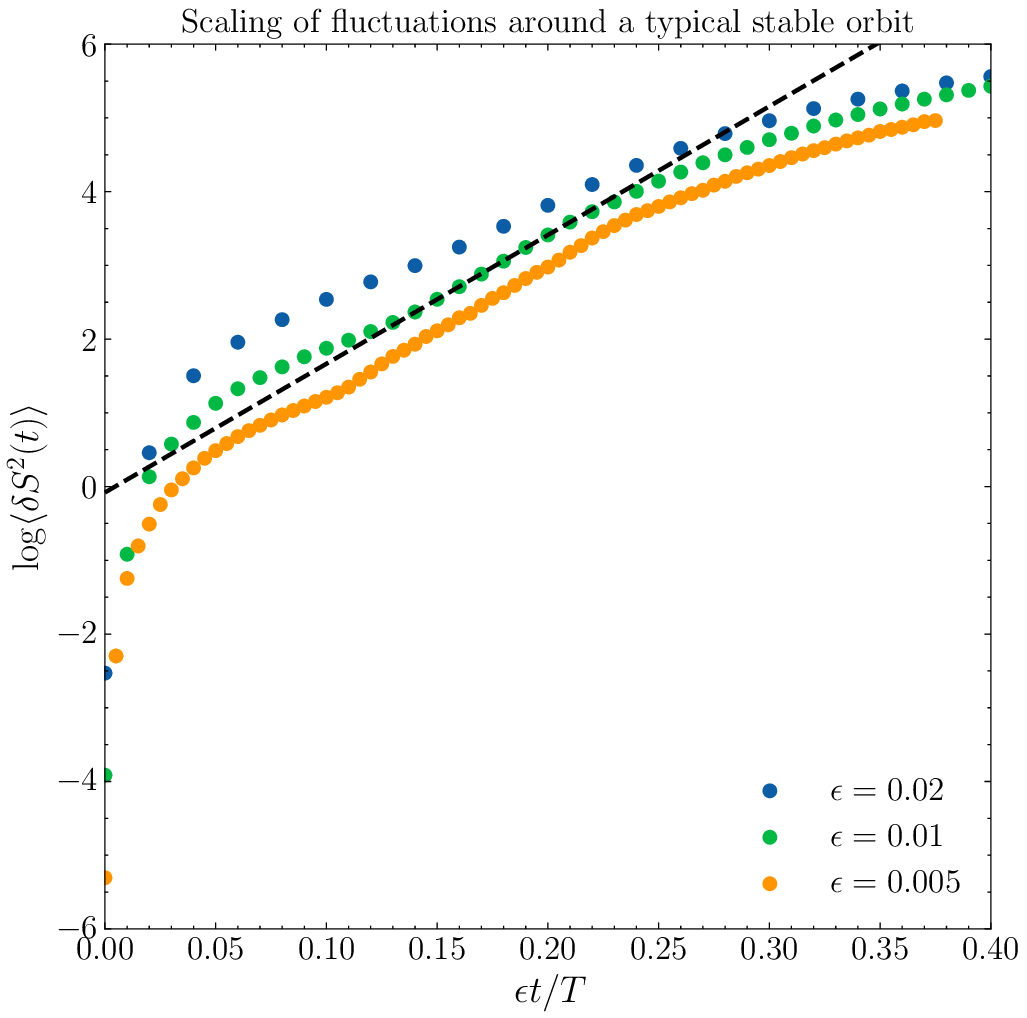}
    \caption{Scaling of relative deviations of a typical stable $\Sigma$-state with different $\epsilon:= \sqrt{\langle \delta s_i^2(t/T=0) \rangle}$. Stable states initially grow linearly, with a subsequent crossover to a parametrically slow exponential growth $\sim e^{{\rm const.} \epsilon\, t/T}$ (indicated by the dashed line), until eventually the fluctuations become of order $O(1)$.
    }
    \label{fig:ds_scaling_stable}
\end{figure}

In summary, the  dynamics of a semiclassical wavepacket close to a stable orbit behaves as shown in Fig.~\ref{fig:ds_scaling_stable}. For $t/T \lesssim \sqrt{S}$ linearized dynamics dominate,  resulting in polynomial growth of the deviations $\delta S \sim \sqrt{t}$ from the  reference orbit of the $S\to\infty$ limit. At times $t/T \sim \sqrt{S}$ the non-linear terms start driving an exponential growth with rate $\kappa \sim 1/\sqrt{S}$. This eventually leads to the blow up of fluctuations, with an associated coherence time scaling as 
\bea
T_{\rm dec}\sim 1/\kappa\sim \sqrt{S}.
\eea

\vspace{10pt}
\section{Instability of trajectories close to the $Z_2$ orbit}
\label{app:instabilityclosetoZ2}
Let us derive how unstable close neighbors of the $Z_2$ orbit are (considering orbits at small distance $||\Delta||$ in orbit space). For the stroboscopic map one expects the analytic expansion for the instability indicator
 \begin{equation}
 \label{TrMclosetoNeel}
     {\rm Tr}(M_k/4) = 1+  r(\Delta) k^2 -   s(\Delta) k^4 + O(||\Delta||^2 k^6).
 \end{equation}
 The functions $r,s$ vanish on the $Z_2$ orbit ($\Delta=0$) and will be argued to be even, and thus  $r,s= O(||\Delta||^2)$. Indeed, since forward propagation from an orbital deviation $\Delta$ is conjugate under $R_z$ to backward propagation from $-\Delta$, and since the stroboscopic maps $M_k$ and their inverse have the same spectrum, it follows that the neighboring orbits starting from $\pm \Delta$ have the same eigenvalues and thus the same value of  ${\rm Tr}(M_k)$, implying $r$ and $s$ to be even. The quadratic part of $r(\Delta)$ turns out to have signature $(+1,-1)$, such that the neighborhood of the $Z_2$ orbit splits into four sectors that alternate between stable ($r<0$) and unstable ($r>0$) regions in the space of 2-periodic orbits (represented by the intersection with the $\Sigma$-plane), cf. Fig.~\ref{fig:stability}.
 
From (\ref{TrMclosetoNeel}) it  follows that the maximal deviations from 1 of eigenvalues of the stroboscopic map are generically associated with $k^*=O(1)$, ${\rm Tr}(M_{k^*}/4)-1 \sim ||\Delta||^2$. Accordingly they scale as $|\lambda_{k^*}-1| = O(||\Delta||)= O(\epsilon)$. 

General perturbations from any stable orbit 
deviate locally from the constraint of zero energy density that characterizes strictly periodic orbits. Those configurations may similarly be expected to show exponentially diverging dynamics with rates $\sim \epsilon$. Therefore a parametrically slow, exponential growth is generically expected to eventually take over the power law growth starting in the vicinity of linearly stable orbits.

\appendix


\end{document}